\documentclass[runningheads]{llncs}
\usepackage{graphicx}
\usepackage{lipsum,subcaption}

\usepackage{hyperref}
\usepackage{amsmath,amsfonts,amssymb}
\usepackage{caption}
\usepackage{subcaption}
\usepackage{float}
\usepackage{multirow}
\usepackage{pifont}
\usepackage{threeparttable}
\usepackage{tablefootnote}
\usepackage{hyperref}
\usepackage[linesnumbered,boxed,ruled]{algorithm2e}
\usepackage{comment}
\usepackage[super]{nth}
\usepackage{balance}
\newcommand*\samethanks[1][\value{footnote}]{\footnotemark[#1]}

\begin{document}
\title{LinkLouvain: Link-Aware A/B Testing and Its Application on Online Marketing Campaign}
\titlerunning{LinkLouvain}
%
\author{Tianchi Cai\thanks{These authors contributed equally.} \and
Daxi Cheng\samethanks \and
Chen Liang\samethanks \and
Ziqi Liu \and
Lihong Gu \and
Huizhi Xie  \and
Zhiqiang Zhang   \and
Xiaodong Zeng  \and
Jinjie Gu
}
\authorrunning{T. Cai et al.}
%
\institute{Ant Financial Services Group, China \\
\email{\{tianchi.ctc,daxi.cdx\}@antfin.com,hi@liangchen.email}}

%
\maketitle              
\begin{abstract}
A lot of online marketing campaigns aim to promote user interaction. The average treatment effect (ATE) of campaign strategies need to be monitored throughout the campaign. A/B testing is usually conducted for such needs, whereas the existence of user interaction can introduce interference to normal A/B testing. With the help of link prediction, we design a network A/B testing method LinkLouvain to minimize graph interference and it gives an accurate and sound estimate of the campaign's ATE. In this paper, we analyze the network A/B testing problem under a real-world online marketing campaign, describe our proposed LinkLouvain method, and evaluate it on real-world data. Our method achieves significant performance compared with others and is deployed in the online marketing campaign.

\end{abstract}

\section{Introduction}

\begin{figure}[tbp] 
\begin{center}
\includegraphics[width=\textwidth]{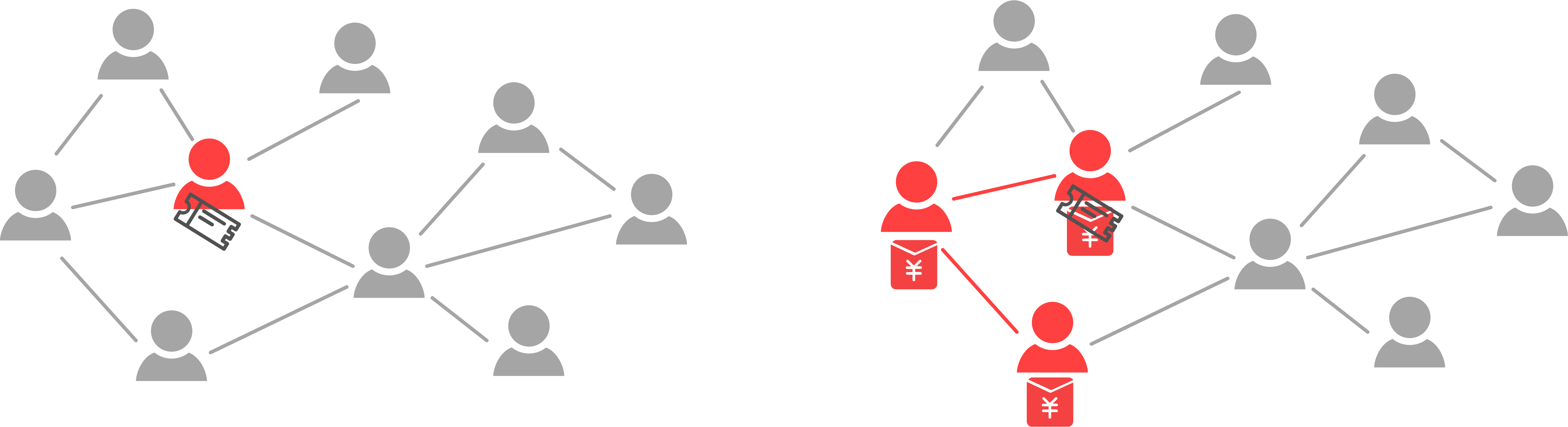}
\end{center}
\caption{Visualization of our online marketing campaign. Coupons are handed out to users (colored red in left figure), and users can invite their friends to join this marketing campaign, and they all receive a cash coupon (colored red in right figure).} \label{fig:line1}
\end{figure}

Recently, Alipay launched an online marketing campaign that encourages users to invite others to join the campaign, so they can all receive discounts or cash rewards. 
Such user interaction-promoting services (IPS) are common to increase user engagement. 
Various strategies are developed for this campaign, and designing an A/B testing solution to quantify their average treatment effects is crucial. 
However, normal A/B testing solutions for IPS are improper because edges (user invitations) exist between different test groups and introduce bias to ATE; A/B testing addressing such interference is called network A/B testing.
Under a thorough analysis of real-world graphs, we develop a graph clustering method LinkLouvain for network A/B testing and deploy it in the online marketing campaign. LinkLouvain has the following strengths: 
\begin{enumerate}
    \item Scalability. It conducts on graphs of billions of nodes and tens of billions of edges in 10 hours.
    \item Simplicity. It is a static method that runs only once before the online marketing campaign. There is no need for additional streaming support.
    \item Effectiveness. It reduces network interference and reduces the heterogeneity of test groups throughout the campaign lifecycle (7 days). We develop two metrics {\itshape estimator bias} and {\itshape estimator variance} to measure the network interference and heterogeneity, respectively. Results show LinkLouvain outperforms others. 
\end{enumerate}

\subsection{Interaction-Promoting Services (IPS)}

For consumer-facing online products, encouraging user interactions is a common practice to increase user engagement. 
Some examples are `People You May Know' on Facebook, `Connections You May Know' on LinkedIn, and online marketing campaigns where coupons can be shared with others on Alipay. 
Such services, referred to as interaction-promoting services (IPS), are designed to encourage user interaction, and therefore benefit user engagement of the product. 

All users and their interactions on the Alipay platform construct a real-world thorough social network with billions of nodes and tens of billions of edges. Users (nodes) and user invitations (edges) in our online marketing campaign form a subgraph of this thorough social network.
Engaging nodes and edges increase throughout the campaign, and this time-evolving graph is always a subgraph of the social network.

In our paper, we analyze the growth properties of the network and the interference patterns from the campaign for a sound understanding of the following network A/B testing problem.

\subsection{Network A/B Testing}

\begin{figure}[tbp] 
\begin{center}
\includegraphics[width=\textwidth]{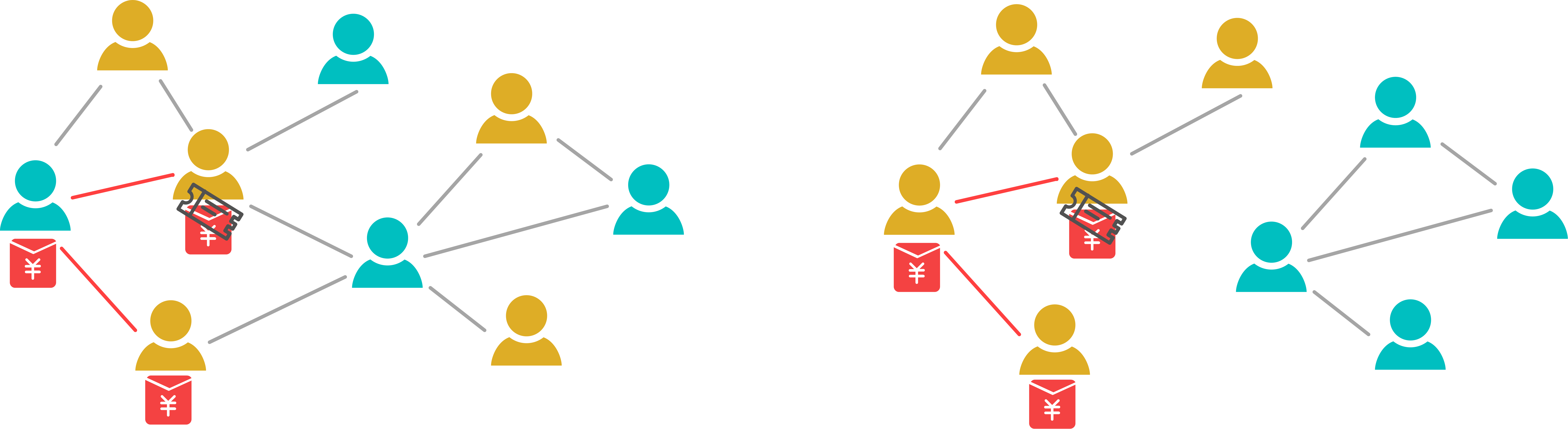}
\end{center}
\caption{Visualization of estimation bias in different A/B testing scheme. \textit{Left}: In user-level randomization, users are randomly selected for the treatment (colored yellow) and the control group (colored cyan). However, the online marketing campaign in the treatment group may affect users in control group. In this case, both the treatment group and the control group has the same number of invited users, and the A/B testing misleadingly concludes that the treatment does not make any difference. \textit{Right} Network A/B testing clusters users with interference together, and the cluster-level metrics show that the treatment group has more invited users.} \label{fig:line2} 
\end{figure}

For IPS, users who do not receive new services may still be affected through interactions with those who do receive new services. 
It introduces interference for user-level A/B testing; thus, a direct estimation of the ATE is no longer unbiased. 
Network A/B testing solutions are of great interest.

There are mainly two approaches to conduct an unbiased estimation of ATE under network interference. The first is afterward correction. 
For example, \cite{gui2015network} assumes the interference is linear-additive, estimates the exposure probability, and weighs the estimation accordingly. The performance of this type of approach relies on making the right decision for the form of interference. We analyze the interference in a real social network, and in our case, however, the linear-additive model is over-simplified and a panacea solution is missing.


The other approach is to perform randomization at the cluster-level. That is, clusters of users, instead of users themselves, are used as randomization units. This approach assumes no/low interference between clusters. Our method falls into this category.

\subsection{Graph Clustering}
Many clustering algorithms have been studied to reduce the interference between their resulting clusters. \cite{ugander2013graph} proposes a clustering algorithm $r$-net. Label propagation and modularity maximization algorithms are also studied in \cite{gui2015network}, and it suggests modularity maximization outperforms the other. However, these approaches usually assume their graphs are restricted-growth graphs (formally defined later) to perform better, which is hard to meet through our analysis of real social networks. Later, we'll introduce our LinkLouvain approach built on Louvain \footnote{A fast and parallel approximation for modularity maximization.}.




We also consider graph partitioning methods to generate balanced test groups with minimal edges between groups. Dynamic graphs at scale impose great challenges for graph partitioning. Most existing algorithms can not scale to billions of nodes. Graph theory based algorithms aiming to solve the optimal min-cut graph partitioning task have been proven NP-hard. Classical graph partitioning methods such as Metis \cite{karypis1998fast} also have high computational complexity. To handle rapidly evolving graphs, classical methods are not favorable for efficiency issues and dynamic graph partitioning algorithms \cite{li2020dynamic,nicoara2015hermes} are proposed by constantly updating labels and graph structure changes that require additional streaming support. 

Our method also focuses on rapidly evolving graphs, however, in a more static manner. Unlike other static methods \cite{stanton2012streaming,tsourakakis2014fennel} that run periodically to obtain continuous partitioning results, we make a guess on the graph structure in the future (e.g. in a week) and partition the predicted graph for only once. In the beginning, we obtain an `omniscient' view of all users and all possible interactions between them. Also, we have a campaign graph in the early stage campaign. Then we predict possible edges with graph neural networks (GNNs) to gain knowledge of a future snapshot of the campaign graph. The current snapshot and future snapshot are formed by invitations in the campaign, while the omniscient graph is irrelevant to specific applications. The predicted edges form a guess of the future snapshot, and it's then clustered by efficient graph clustering methods with linearithmic time such as Louvain. Finally, the clusters are randomly merged to $p$ desired test groups for A/B testing.

\section{Preliminaries}
\subsection{Problem Formulation}
In this paper, we are interested in the task of network A/B testing. More specifically, we aim at estimating a precise and sound ATE. 
Estimating ATE when launching or updating IPS, however, is non-trivial. 
In the absence of interactions, user-level A/B testing is commonly used to estimate potential effects \cite{kohavi2009controlled}. The estimation is unbiased if the {\itshape Stable Unit Treatment Value Assumption} (SUTVA) holds. This assumption requires the response of a unit (in this case, a user) to be invariant to treatments assigned to other units \cite{cox1958planning}. With this assumption, the {\itshape average treatment effect} (ATE) of a new service can be defined as
\begin{displaymath}
    ATE=\frac{1}{N}\sum_i y_1(i)-y_0(i),
\end{displaymath} 
where $y_0(i)$ is the outcome for user $i$ if not treated and $y_1(i)$ is the outcome for user $i$ if treated. $N$ represents the number of users.

However, the ground-truth ATE in real-world network A/B testing is impossible to obtain. Our work designs an estimator of ATE in the presence of network interference by splitting graph to clusters. The estimator is formulated as 
\begin{displaymath}
    \hat{ATE}=\frac{1}{M}\sum_i \sum_j y_1(q_j^i)-\frac{1}{N}\sum_i \sum_j y_0(c_j^i),
\end{displaymath} 
where $q_j^i$ is $i$-th user in $j$-th cluster of the treatment group $Q$ and $c_j^i$ is $i$-th user in $j$-th cluster of the control group $C$. $M$ and $N$ represent numbers of users in $Q$ and $C$, respectively.

Our goal is to design an ATE estimator that minimizes the estimation bias and variance. Therefore, the estimated ATE can guide business decisions.

\subsection{Two Graphs of Interest}

In our online marketing campaign, we have access to two graphs: a stable social graph $G=(V,E)$ and a time-evolving label graph $L = (V_L, E_L)$. 
We collect the social graph $G$ containing all users of Alipay as nodes $V$ and their historical interactions as edges $E$. It contains billions of nodes and tens of billions of edges and lays the foundation for predicting users' future interactions. 

Additionally, as the new online marketing campaign goes on, we collect a label graph $L$, where users who participate in the online marketing campaign form node set $V_L$ and user invitations form edge set $E_L$. $L^0$ and $L^T$ represent the label graph in its early stage and by the end of the campaign of lifecycle $T$, respectively. 
It is called a label graph since the interaction data provides a strong hint for the form of interference. Previously, labeled data is less discussed because users already participated cannot join a new round of A/B testing. The novelty of LinkLouvain is that it uses link prediction to generalize the form of interference from this label graph to all users in the social graph $G$, and predicts an ``estimator bias" for all edges $V$.

Various properties of the two collected graphs are analyzed in section~\ref{sec:property}.

\section{The Proposed Framework}

\begin{figure}[h]
  \centering
  \includegraphics[scale=0.49]{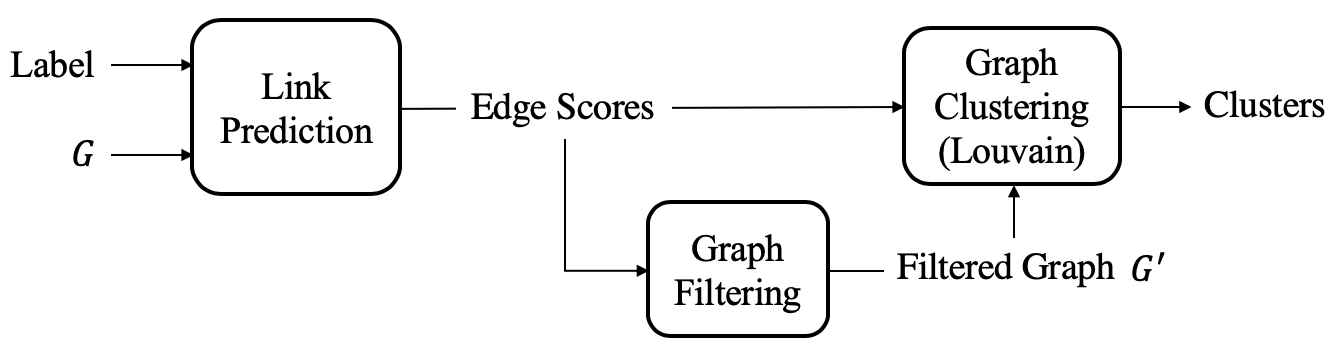}
  \caption{Processing pipeline of the proposed LinkLouvain framework.}
  \label{fig:process}
\end{figure}

To cluster a rapidly evolving graph, we train a GNN based link prediction model to predict possible edges in the evolving graph. Then we apply a traditional graph clustering algorithm such as Louvain to split the graph into small clusters. To use these clusters in A/B testing, we randomly combine them into desired $p$ test groups. The procedure is shown in Figure~\ref{fig:process}. {\itshape Label} comes from the edges (positive labels) in the current campaign graph and non-edges (negative labels) that exist only in the social graph $G$.

\subsection{GNN based Link Prediction Models}

GNNs are a set of deep learning architectures that aggregate information from nodes' neighbors using neural networks. Deeper layers aggregate more distant neighbors, and the $k$th layer embedding of node $v$ is \begin{displaymath}
  \mathbf{h}_v^k = \sigma (\mathbf{W}_k \cdot  \mathrm{AGG}(\mathbf{h}_u^{k-1},\forall u \in \mathcal{N}{(v)} \cup \{v\})) 
\end{displaymath} where the initial embedding $\mathbf{h}_v^0=\mathbf{x}_v$ is its node feature vector, $\sigma$ is a non-linear function, and $\mathrm{AGG}$ is an aggregation function that differs in GNN architectures.

\begin{figure}[h]
  \centering
  \includegraphics[scale=0.28]{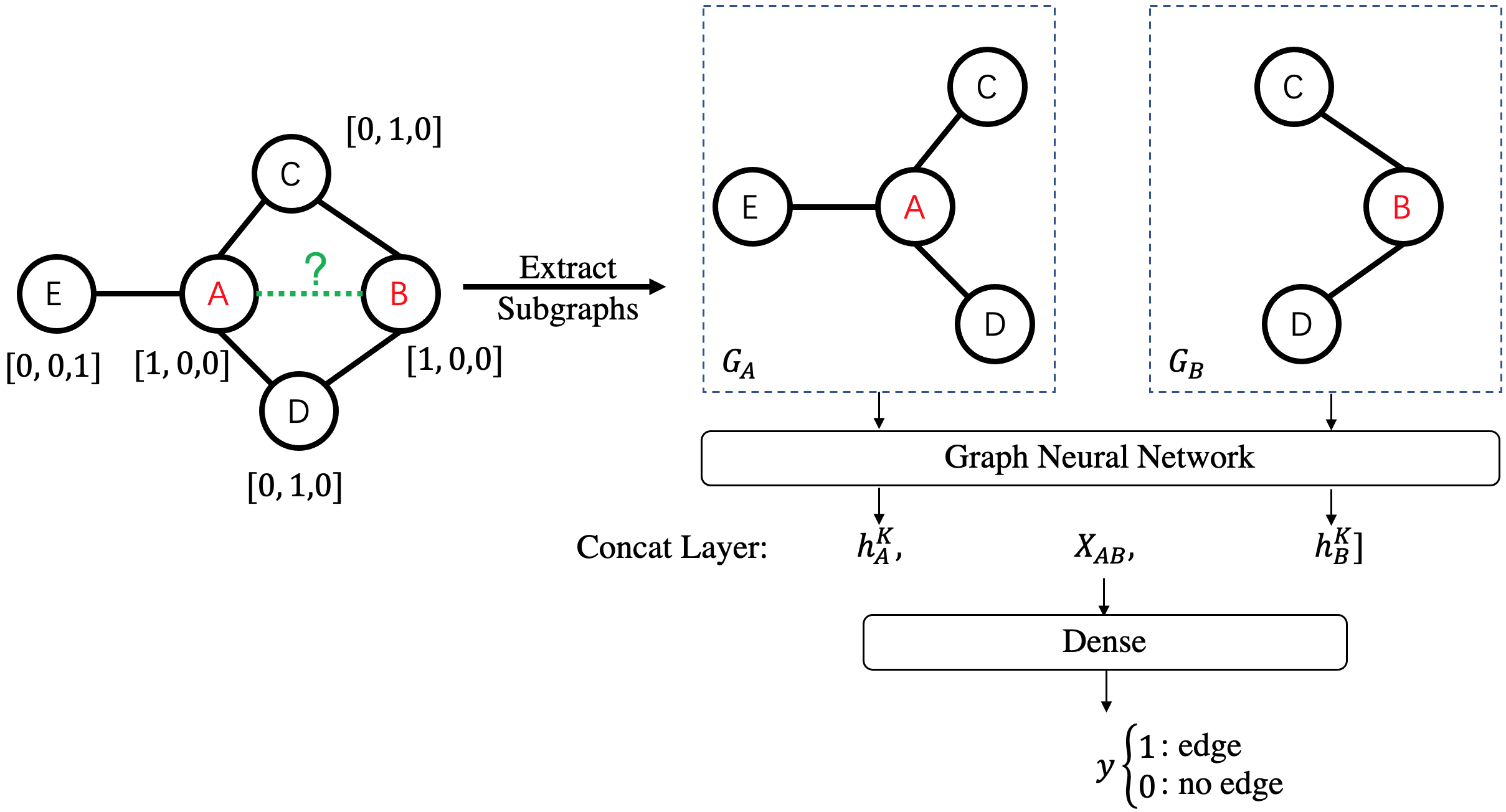}
  \caption{Model architecture for link prediction. $G$: 1-hop neighborhood of a node; $X$: edge features; $h$: GNN embeddings; one-hot vectors: node labeling.}
  \label{fig:gnn}
\end{figure}

Figure~\ref{fig:gnn} shows a naive GNN based link prediction algorithm with a twin-tower architecture. Each target node of an edge aggregates its own neighbors for $K$ times. After aggregation of $K$-hop neighbors, the final embeddings $\mathbf{h}_A^K$ and $\mathbf{h}_B^K$ of two target nodes $A$ and $B$ are concatenated and fed to the final dense layer.

Moreover, we add structural features called node labeling \cite{zhang2018link} to naive link prediction. Node labeling assigns a one-hot vector to each node in the $K$-hop neighborhood of two target nodes $A$ and $B$. It marks nodes' different roles in this neighborhood. For example, the left graph in Figure~\ref{fig:gnn} has 5 nodes in $A$ and $B$'s 1-hop neighborhood. There are three roles in the neighborhood: $A$ and $B$ are target nodes; $C$ and $D$ are nodes connecting both target nodes; $E$ is a node that only connects to one target. 
The node labeling vector is appended to each node's original feature vector and tells GNN its relative location around the edge to be predicted. It helps GNN to have more accurate predictions on link existence.

\subsubsection{Comparison of Link Prediction Models in Online Marketing Campaign}

In the early stage of the online marketing campaign, we collect and sample user interactions as positive training samples and non-invitation relations as negative training samples. All training samples exist in our social graph $G$. There are 1.5 million positive edges and 1.5 million negative edges. Each edge has 128 features representing user interaction history.
We compare the following models for the link prediction task: 
\begin{itemize}
\item DNN: a dense neural network of five layers with layer size $[512, 256,$ $128, 64, 16]$.
\item NG-LP: a naive GNN link prediction method with 2-hop neighbors ($K=2$) and embedding size 64.
\item NL-LP: a node labeling link prediction method with 2-hop neighbors ($K=2$) and embedding size 64.
\end{itemize}

The main results are summarized in Table~\ref{tab:gnn}. F1\footnote{\url{https://en.wikipedia.org/wiki/F1_score}}, KS\footnote{\url{https://en.wikipedia.org/wiki/Kolmogorov-Smirnov_test}}, and AUC \footnote{\url{https://en.wikipedia.org/wiki/Receiver_operating_characteristic}} are widely used binary classification metrics. NL-LP performs the best by taking structural information into account.

\begin{table}
\centering
\caption{Link prediction task comparison.}
\label{tab:gnn}
    \begin{tabular}{l|c|c|c}    
        
         & DNN & NG-LP & NL-LP \\
        \hline
        F1 & 0.88 & 0.89 & {\bf 0.91} \\
        KS & 0.74 & 0.79 & {\bf 0.84} \\
        AUC & 0.91 & 0.92 & {\bf 0.96} \\
        \end{tabular}
\end{table}

\subsection{Graph Filtering}

The output scores on edges represent possibilities of future online interactions. We filter out less possible edges and set the prediction score as edge weight. Graph filtering is crucial for a billion-node graph and the reasons are two-fold:

\begin{itemize}
    \item Computation resources are limited for graphs of such size.
    \item Clustering algorithms like Louvain tend to generate unbalanced clusters when handling densely connected graphs. They undermine A/B testing performance heavily. Removing unnecessary edges help prevent long tails of resulting clusters.
\end{itemize}

 However, if we set the threshold ($\gamma$) to abandon or keep an edge too high, we could drop too many possible edges. This introduces great bias on ATE estimates. We choose $\gamma$ considering the trade-off between efficiency and effectiveness.

 In the online marketing campaign, we set the threshold to be 0.5, and clustering the remaining graph costs 0.6 hours.

\subsection{Graph Clustering}

To generate clusters of users as randomization units, we use Louvain to cluster the filtered graph $G^\prime$ . Clustering algorithms are well-discussed. In \cite{gui2015network}, researchers investigate several distributed clustering algorithms, such as label propagation and Louvain. Their result shows that Louvain performs better in preserving more intra-cluster edges and reducing network interference. Experiment results in the next section (Table~\ref{tab:summary}) also support this conclusion. 

The resulting clusters are finally randomly merged into partitions of the desired size $p$. These are the $p$ test groups in A/B testing.
    

\section{Application on Online Marketing Campaign}

\begin{figure*}[!tbp]
     \centering
     \begin{subfigure}[b]{0.475\linewidth}
         \centering
         \includegraphics[width=\textwidth]{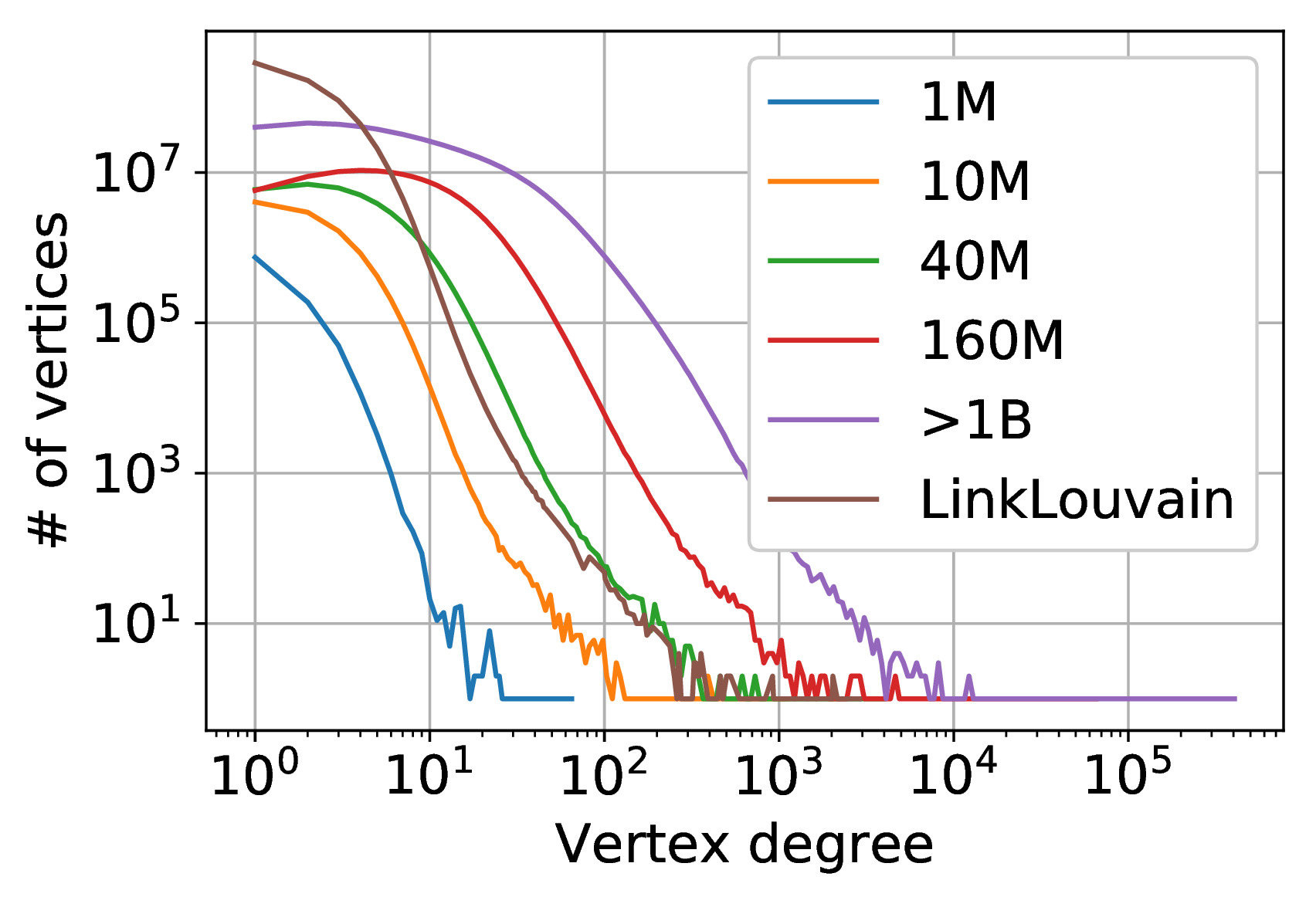}
         \label{graph:degree}
     \end{subfigure}
     \begin{subfigure}[b]{0.50\linewidth}
         \centering
         \includegraphics[width=\textwidth]{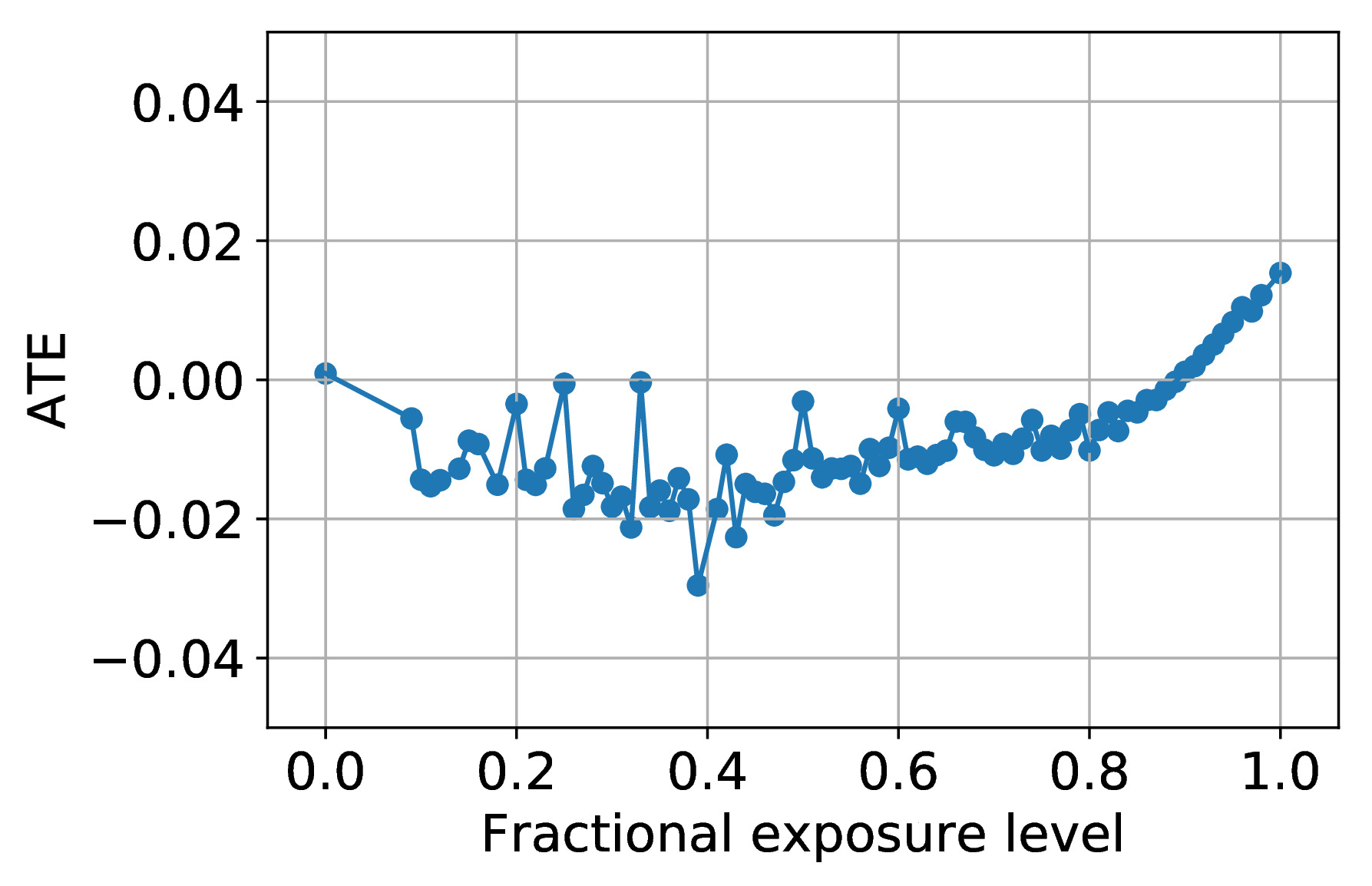}
         \label{graph:neighborhood}
     \end{subfigure}
     
     \caption{\textit{Left}: The vertex degree distribution of our real social network $G$ at different growth levels, as well as $G$ after graph filtering by LinkLouvain. \textit{Right}: Since the treatment effect of a user depends on his/her neighborhood’s treatment status, there exists interference. Moreover, this influence is non-linear to the fractional exposure level, and cannot be corrected afterward easily.}
     \label{fig:3graphs}
\end{figure*}

\subsection{Patterns of Our Real-World Graphs}
\label{sec:property}

Though $G$ is a large social network that does not change frequently, the size of $L$ grows quickly as users joining our campaign. Therefore we can analyze the growth property of our social network. As the number of nodes in $L$ reaches 1, 10, 40, 160 millions, we construct a subgraph of $G$ with all nodes in $L$, and keep all edges between these nodes. Hence we can analyze the growth property of our social network retrospectively. We compare the graph properties of these four subgraphs of $G$, as well as the full graph $G$, which contains more than 1 billion nodes. 

\subsubsection{Maximum Degree Growth is Unbounded} 
In Figure \ref{fig:3graphs} \textit{Left}, we compare the degree distribution of $G$ at different growth levels. We find that as the network grows larger, customers build more connections with each other, and the degree distribution shifts right. The long right tails of all five series suggest that the degree of this social network has a right-skewed distribution regardless of the network size. Moreover, diverged from bounded maximum degree assumptions \cite{ugander2013graph}, the maximum degree grows almost linearly to the number of nodes in the graph, and thus, unbounded.

In Figure \ref{fig:3graphs} \textit{Left}, We also plot the degree distribution of the full graph $G$ after graph filtering by LinkLouvain. It is clear that the degree distribution is less skewed compared to the original distribution of the full social graph $G$ (labeled "$>$1B"). The intuition behind is that not all edges in $G$ have the same influence on our online marketing campaign. Therefore we can eliminate many edges that are not likely to have interactions with LinkLouvain and hence reduce interference in our cluster-level randomization scheme.




\subsubsection{Network Interference}
We examine network interference patterns on our social network by estimating the ATE on different {\itshape neighborhood fractional exposure level} \cite{gui2015network} (share of neighbors that are in the treatment group) to see if there is any pattern. We divide users into subgroups according to their different fractional exposure levels and plot estimate ATEs with respect to each group as a curve in figure \ref{fig:3graphs} \textit{Right}. We can draw two main conclusions. First, the treatment effect of a user depends on his/her neighborhood's treatment status, which means that the interference exists. Second, the interference does not follow a linear-additive pattern; in other words, the ATE is not linear with the fractional exposure level.

This explains the difficulty of using the afterward correction approach: there is no universal assumption for the form of interference suitable for all cases. The true form of interference might be complicated, and the linear-additive assumption might be over-simplified.

\subsection{Metrics}
\label{sec:metrics}
Lower estimator bias and variance indicate more accurate and sound estimations. Here we introduce how to measure them.

\subsubsection{Estimator Bias}

is measured by the degree of network interference between test groups. Clusters of users are randomly merged to $p$ desired A/B test groups. Edges of graph $L^T$ exist between test groups, and their interference is denoted as $I=\frac{|E^-|}{|E_L^T|}$, where $E_L^T$ is the set of all edges (invitations in the campaign) in graph $L^T$, and $E^-$ is the set of edges in graph $L^T$ connecting nodes across test groups.

\subsubsection{Estimator Variance} represents the statistical power of designed estimators. To get higher statistical power, our estimator should generate clusters where the ATE metric of the clusters has a lower variance, which means online experiments are more sensitive. We use the following formula from \cite{deng2018applying} to calculate the variance of clusters in an estimator,
\begin{displaymath}
    \operatorname{Var}(\overline{Y})\approx\frac{1}{K\mu_{N}^{2}}\left(\sigma_{S}^{2}-2\frac{\mu_{S}}{\mu_{N}}\sigma_{SN}+\frac{\mu_{S}^{2}}{\mu_{N}^{2}}\sigma_{N}^{2}\right),
\end{displaymath}where $\bar{Y}$ is the total estimated conversion rate in this A/B testing group. $K$ is the number of clusters in the group. $S$ and $N$ are the random variables of the sum of individual conversion and individual number respectively. $\mu$ and $\sigma$ calculate the mean and variance/covariance of the corresponding random variables. We evaluate the metric variance with the same group size (1\% of the total traffic).

\subsection{Methods for Comparison}
The methods in our comparative evaluation are as follows.
\begin{itemize}
    \item Geo: the classical strategy to cluster users by their geographic locations.
    \item LinkLouvain: our proposed method with graph filtering threshold $\gamma$.
    \item Louvain: an ablation study that removes the link prediction stage and the graph filtering stage.
    \item HRLouvain: an ablation study that replaces the link prediction stage and the graph filtering stage by removing hotspots (nodes with more than $\theta$ neighbors).
    \item LinkLouvain-UW: an ablation study that replaces link prediction edge weight by 1 in our proposed method.
    \item LinkLabel: an ablation study with Louvain replaced by label propagation for graph clustering.
\end{itemize}

\subsection{Evaluation}

Table~\ref{tab:summary} summarizes the evaluation results of all the methods on our campaign. Metrics include estimator bias and estimator variance described in Section~\ref{sec:metrics} as well as computation time. The number of clusters is also summarized for reference.

\begin{table}
\caption{Evaluation summary. (Louvain runs for more than one day and drains computational resources. Its results are not available.)}
\label{tab:summary}
\centering
    \begin{tabular}{l|c|c|c|c}    
        Methods & \# of clusters & $I$ & $Var(\bar{Y})$ & Time \\        
            &  &  (\%) & $(10^{-8})$ & (h)  \\
        \hline
        Geo  & 346 & {\bf 52} & 47890 & 0.2 \\    
        \hline
        LinkLouvain, $\gamma=0.2$ & 206M & {\bf 50} & 1.17 & 12.7 \\
        \hline
        LinkLouvain, $\gamma=0.3$ & 248M & {\bf 49} & 1.15 & 9.8   \\  
        \hline
        LinkLouvain, $\gamma=0.5$ & 359M & {\bf 52} & {\bf 1.11} & {\bf 5.6} \\ 
        \hline
        Louvain  & - & - & - & $>$24 \\   
        \hline
        HRLouvain, $\theta=40$ & 367M & 82 & {\bf 1.10} & {\bf 5.4} \\
        \hline
        HRLouvain, $\theta=100$ & 145M & 67 & 32.45 & 12.1\\
        \hline
        HRLouvain, $\theta=200$ & 98M & 66 & 232.62 & 12.6\\
        \hline
        LinkLouvain-UW  & 442M & 64 & {\bf 1.07} & 6.1 \\
        \hline
        LinkLabel & 351M & 67 & 1.37 & 10.2 \\
        
        \end{tabular}

\end{table}

In general, LinkLouvain shows effectiveness in delivering precise and sound estimates and efficiency to run within 6 hours.

\subsubsection{Consistency}
We compare three sets of threshold $\gamma$ (0.2, 0.3, and 0.5) for LinkLouvain, and their key metrics are consistent. It leads to an easier tuning process during experiments.

\subsubsection{Computational Performance} We run clustering methods with 40 workers on GRAPE~\cite{fan2018parallelizing}. Table~\ref{tab:summary} summarizes the computation time, and LinkLouvain with $\gamma=0.5$ and HRLouvain with $\theta=40$ are the most efficient in graph based methods.

\subsubsection{Comparison with Geo-based Methods} A popular way to run A/B testing in online services is to use geographic regions as randomization units. It serves as a practical baseline for comparison. It is easy to use since it only requires locations of user queries. We compare our method with geo-based partitioning, and the results show we achieve much lower variance compared to this popular approach.

\subsection{Online Results}
The online campaign run for 7 days and our LinkLouvain ($\gamma=0.5$) method was deployed to give estimates of ATE of different campaign strategies such as giving discount coupons or cash coupons. The ATE is the average payment made by users who receive coupons, and the ATE estimate of the best strategy is 1.05 times better than baseline (giving everyone a small amount of cash) without increasing the campaign budget. The A/B test was run on 2\% users in the campaign, and after monitoring strategies for a day, the best coupon-distributing strategy was applied to 100\% users. The performance of the campaign exceeds expectations with the help of LinkLouvain.

\section{Conclusion}

In this paper, we discuss network A/B testing motivated by interaction-promoting services. We analyze this problem in a real social graph and our label graph and develop LinkLouvain to address network A/B testing. The proposed approach is computationally efficient and achieves the preferable balance between estimator bias and estimator variance with the help of link prediction. It is deployed on a real marketing campaign and gives accurate and sound estimates of ATEs.

\bibliographystyle{acm}
\bibliography{sample-base}

\begin{thebibliography}{10}

\bibitem{cox1958planning}
{\sc Cox, D.~R., and Cox, D.~R.}
\newblock {\em Planning of experiments}, vol.~20.
\newblock Wiley New York, 1958.

\bibitem{deng2018applying}
{\sc Deng, A., Knoblich, U., and Lu, J.}
\newblock Applying the delta method in metric analytics: A practical guide with
  novel ideas.
\newblock In {\em Proceedings of the 24th ACM SIGKDD International Conference
  on Knowledge Discovery \& Data Mining\/} (2018), pp.~233--242.

\bibitem{fan2018parallelizing}
{\sc Fan, W., Yu, W., Xu, J., Zhou, J., Luo, X., Yin, Q., Lu, P., Cao, Y., and
  Xu, R.}
\newblock Parallelizing sequential graph computations.
\newblock {\em ACM Transactions on Database Systems (TODS) 43}, 4 (2018),
  1--39.

\bibitem{gui2015network}
{\sc Gui, H., Xu, Y., Bhasin, A., and Han, J.}
\newblock Network a/b testing: From sampling to estimation.
\newblock In {\em Proceedings of the 24th International Conference on World
  Wide Web\/} (2015), pp.~399--409.

\bibitem{karypis1998fast}
{\sc Karypis, G., and Kumar, V.}
\newblock A fast and high quality multilevel scheme for partitioning irregular
  graphs.
\newblock {\em SIAM Journal on scientific Computing 20}, 1 (1998), 359--392.

\bibitem{kohavi2009controlled}
{\sc Kohavi, R., Longbotham, R., Sommerfield, D., and Henne, R.~M.}
\newblock Controlled experiments on the web: survey and practical guide.
\newblock {\em Data mining and knowledge discovery 18}, 1 (2009), 140--181.

\bibitem{li2020dynamic}
{\sc Li, H., Yuan, H., Huang, J., Cui, J., and Yoo, J.}
\newblock Dynamic graph repartitioning: From single vertex to vertex group.
\newblock In {\em International Conference on Database Systems for Advanced
  Applications\/} (2020), Springer, pp.~482--497.

\bibitem{nicoara2015hermes}
{\sc Nicoara, D., Kamali, S., Daudjee, K., and Chen, L.}
\newblock Hermes: Dynamic partitioning for distributed social network graph
  databases.
\newblock In {\em EDBT\/} (2015), pp.~25--36.

\bibitem{stanton2012streaming}
{\sc Stanton, I., and Kliot, G.}
\newblock Streaming graph partitioning for large distributed graphs.
\newblock In {\em Proceedings of the 18th ACM SIGKDD international conference
  on Knowledge discovery and data mining\/} (2012), pp.~1222--1230.

\bibitem{tsourakakis2014fennel}
{\sc Tsourakakis, C., Gkantsidis, C., Radunovic, B., and Vojnovic, M.}
\newblock Fennel: Streaming graph partitioning for massive scale graphs.
\newblock In {\em Proceedings of the 7th ACM international conference on Web
  search and data mining\/} (2014), pp.~333--342.

\bibitem{ugander2013graph}
{\sc Ugander, J., Karrer, B., Backstrom, L., and Kleinberg, J.}
\newblock Graph cluster randomization: Network exposure to multiple universes.
\newblock In {\em Proceedings of the 19th ACM SIGKDD international conference
  on Knowledge discovery and data mining\/} (2013), pp.~329--337.

\bibitem{zhang2018link}
{\sc Zhang, M., and Chen, Y.}
\newblock Link prediction based on graph neural networks.
\newblock In {\em Advances in Neural Information Processing Systems\/} (2018),
  pp.~5165--5175.

\end{thebibliography}
\end{document}